\documentclass[aps,prl,twocolumn,superscriptaddress]{revtex4-2}
\usepackage{bbm}
\usepackage{graphicx}
\usepackage{dcolumn}
\usepackage{bm}
\usepackage{subfigure}
\usepackage{amsmath}
\usepackage{feynmf}
\usepackage{hyperref}
\usepackage{amssymb}
\usepackage{attachfile}
\usepackage{braket}
\usepackage{esint}
\usepackage[dvipsnames,svgnames]{xcolor}

\definecolor{midnightblue}{rgb}{0.1, 0.1, 0.44}
\definecolor{navy}{rgb}{0.0, 0.0, 0.5}
\definecolor{myperfectblue}{rgb}{0.2, 0.3, 0.9}

\begin{document}

	\title{Electron Dynamics Reconstruction and Nontrivial Transport by Acoustic Waves}
	
	\author{Zi-Qian Zhou}
	\affiliation{International Center for Quantum Materials, School of Physics, Peking University, Beijing 100871, China}
	
	\author{Zhi-Fan Zhang}
	\affiliation{Interdisciplinary Center for Theoretical Physics and Information Sciences, Fudan University, Shanghai 200433, China}
	\affiliation{State Key Laboratory of Surface Physics, Fudan University, Shanghai 200433, China}
	
	\author{Cong Xiao}
	\affiliation{Interdisciplinary Center for Theoretical Physics and Information Sciences, Fudan University, Shanghai 200433, China}
	
	\author{Hua Jiang}
	\email{jianghuaphy@fudan.edu.cn}
	\affiliation{Interdisciplinary Center for Theoretical Physics and Information Sciences, Fudan University, Shanghai 200433, China}
	\affiliation{State Key Laboratory of Surface Physics, Fudan University, Shanghai 200433, China}
	
	\author{X. C. Xie}
	\email{xcxie@pku.edu.cn}
	\affiliation{International Center for Quantum Materials, School of Physics, Peking University, Beijing 100871, China}
	\affiliation{Interdisciplinary Center for Theoretical Physics and Information Sciences, Fudan University, Shanghai 200433, China}
	\affiliation{Hefei National Laboratory, Hefei 230088, China}

	\date{\today}
	
	\begin{abstract}
		
		Surface acoustic waves (SAWs) become a popular driving source in modern condensed matter physics, but most existing theories simplify them as electric fields and ignore the non-uniform Brillouin zone folding effect. We develop a semiclassical framework and reconstruct the electron dynamics by treating SAW as a quasi-periodic potential modulating electronic momentum distribution. This framework naturally explains the experimentally observed DC drag current and predicts acousto-electric Hall effect. The theory further reveals various SAW-driven transport phenomena, emerging anomalous Hall, thermal Hall, and Nernst effects within time-reversal symmetric systems. Illustrated in bilayer graphene and $\mathrm{MX_2}$ (M = Mo, W; X = S, Se, Te), the angular-dependent acousto-electric Hall effect provides an experimental probe for Berry curvature distribution.
		
	\end{abstract}
	
	\maketitle
	
		\begin{figure}[t]
		\centering
		\includegraphics[width=0.9\linewidth]{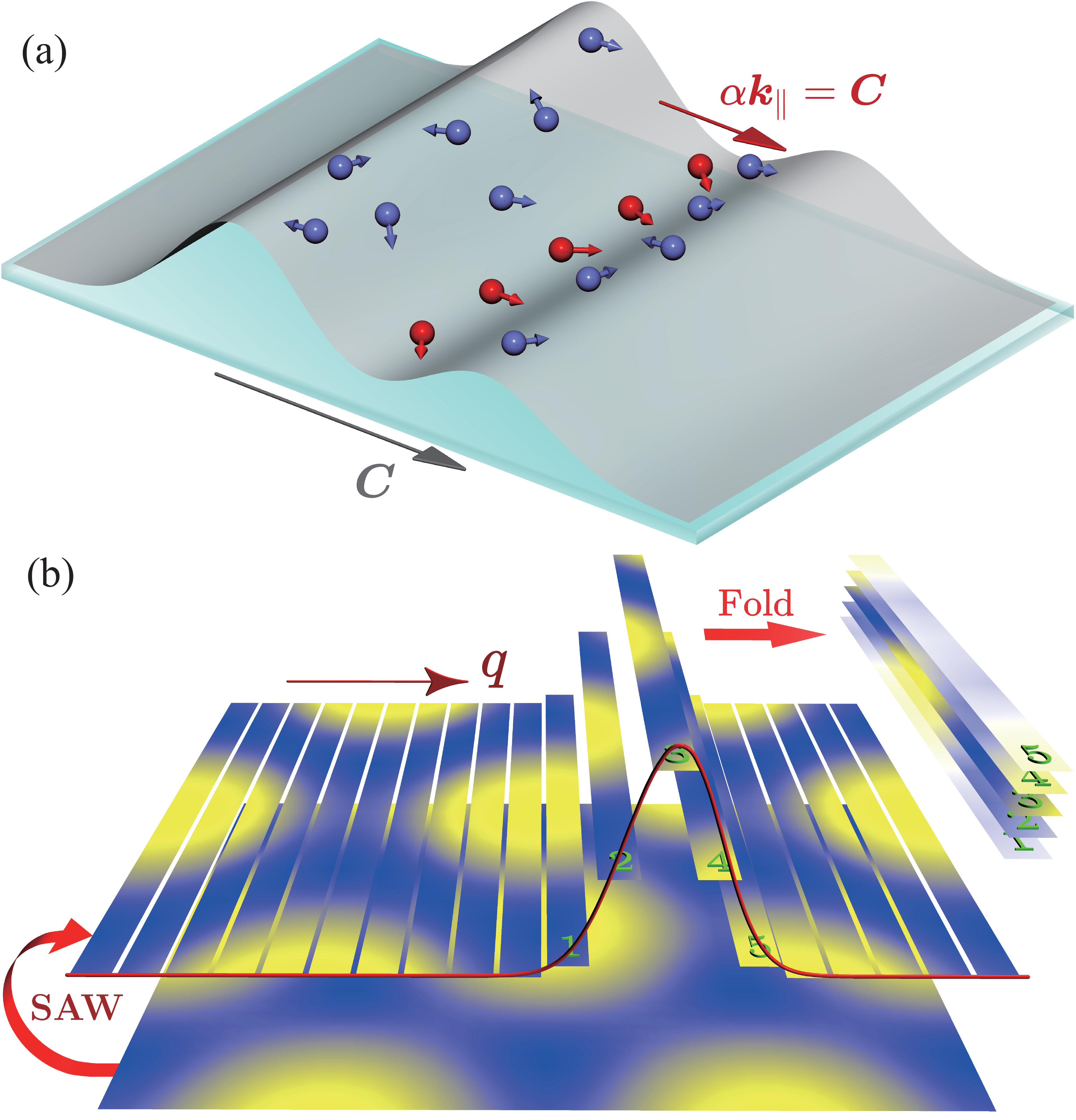}
		\caption{(a) In the SAW system, electrons with momentum satisfying $\alpha\boldsymbol{k}_{\parallel}= \boldsymbol{C}$ become strongly bound in the potential valleys of the SAW-induced electric potential $V(\boldsymbol{x}, t)$. (b) Bottom panel: Brillouin zone and Berry curvature for a system with time reversal symmetry, resulting in Hall conductivity $\sigma_{xy}=0$. Top panel: The Brillouin zone undergoes SAW-induced non-uniform folding. The dominant contribution arises from the region near the specific momentum $\boldsymbol{k}_{\parallel} = \boldsymbol{C}/\alpha$, while contribution of other momenta diminish significantly. Non-zero $\sigma_{xy}$ is obtained.}
		\label{Fig1}
	\end{figure}
	
	{\it Introduction.---}Surface acoustic waves (SAWs) represent one of the most fundamental driving fields in condensed matter physics, with decades of extensive research. As a critical technique for wave manipulation, SAW technology has reached maturity in its theoretical framework, experimental implementation, and technical applications \cite{slobodnik1976surface,seifert1994mechanical,mayer1995surface,PhysRevB.58.7958,Hess2002Surface,de2005modulation, lange2008surface, yeo2014surface,PhysRevLett.134.026304}. Particularly, SAWs have been used to probe and control elementary excitations including electrons, magnons, and excitons in recent years, demonstrating their sustained value to the scientific community \cite{zhang2020surface,gustafsson2014propagating,mcneil2011demand,preciado2015scalable,yokouchi2020creation,fang2023quantum,yuan2021remotely,peng2022long,datta2022spatiotemporally,roshchupkin2022acoustically,mou2024gate,buhler2022chip,nie2023surface,yamamoto2020non}. SAWs have been experimentally demonstrated to induce the acousto-electric effect \cite{PhysRev.89.990}, with SAW driving electrons in a manner comparable to electric fields \cite{doi:10.1021/acs.nanolett.4c02941}. In topological materials, such acousto-dragged electrons should exhibit anomalous velocities, analogous to electric-field-induced counterparts.
	
	Semiclassical electron dynamics plays a pivotal role in condensed matter physics, particularly in linking topological quantities to electron transport \cite{thouless1982quantized, thouless1983quantization, king1993theory, laughlin1983anomalous,berry1984quantal}. The semiclassical dynamics of Bloch electrons, intimately connected with quantum geometric quantities \cite{RN118,sundaram1999wave, RN119, RN123, RN134, RN128}, provides a powerful physical picture for understanding exotic quantum transport phenomena. This framework successfully explains effects such as anomalous charge transport and anomalous thermoelectric response within a unified theoretical approach \cite{PhysRevB.59.14915,PhysRevB.68.045327,PhysRevLett.97.026603,PhysRevLett.126.187001}. Thus, within the semiclassical electron dynamics framework, we reveal the overlooked modulation of electrons in SAW-driven transport.

	

	
	The effects of SAWs on electrons have traditionally been treated as an equivalent electric fields, preserving the semiclassical electron dynamics structure unchanged \cite{Kalameitsev2019Valley, Sonowal2020Acoustoelectric, PhysRevLett.124.126602}. However, in experiments, the amplitude of the effective potential induced by SAW is often non-negligible, causing the SAW to selectively bind electrons whose momenta match the SAW velocity (Fig. \ref{Fig1}(a)). As a result, the folding of the Brillouin zone is no longer an equal-weight superposition but instead adopts a Gaussian-like profile (Fig. \ref{Fig1}(b)). This Gaussian-like folding fundamentally alters the electron dynamics by replacing the Berry curvature to the folded Berry curvature, and introduces additional terms in the equations of motion originating from the non-uniform folding.

	
	
	In this work, we develop a novel semiclassical framework for describing electron dynamics in SAW systems that explicitly accounts for non-uniform Brillouin zone folding. Our theory explains the experimentally observed longitudinal DC current proportional to the SAW decay rate \cite{PhysRevLett.128.256601} and predicts the emergence of a previously unreported acousto-electric Hall effect, even in topologically trivial systems. Moreover, this framework reveals the existence of anomalous Hall effects, thermal Hall effects, and Nernst effects in SAW systems. We demonstrate the angular dependence of acousto-electric Hall effect in both bilayer graphene and monolayer transition metal dichalcogenides (TMDs) as illustrative examples.


	
	

	
	{\it Electron wave packet affected by quasiperiodic field.---} We consider the problem within a piezoelectric material substrate, where SAWs manifest as attenuated traveling waves \cite{Mikoshiba1963Magnetic, PhysRevLett.128.256601, Gu2024Acousto}. One of the SAW modes can be expressed as a rightward-traveling wave,
	\begin{equation}
		\boldsymbol{u}\left(\boldsymbol{x},t\right) = \boldsymbol{s}e^{-\boldsymbol{\mu}\cdot\left(\boldsymbol{x}-\boldsymbol{C}t\right)}\sin\left[\boldsymbol{q}\cdot\left(\boldsymbol{x}-\boldsymbol{C}t\right)\right],
	\end{equation}
	where $\boldsymbol{s}$ represents the amplitude of SAW. $\boldsymbol{\mu}$ is the attenuation factor related to the longitudinal conductance $\sigma_{xx}$ of the material \cite{doi:10.1021/acs.nanolett.4c02941,Rotter1998Giant, Tang2017Ultra,Friess2017Negative,wang2024anomalousacoustocurrentquantumhall,wu2024probingquantumphasesultrahighmobility}. $\boldsymbol{q}$ denotes the wave vector of the SAW, which is aligned with $\boldsymbol{\mu}$. $\boldsymbol{C}$ is the SAW velocity. {All parameters $\boldsymbol{\mu}$, $\boldsymbol{q}$, and $\boldsymbol{C}$ are expressible through the system's intrinsic microscopic parameters and the SAW frequency $\omega$ \cite{SupplementalMaterials}, with values accessible via first-principles calculations or experimental measurements.} Taking into account the displacement of ion cores and the uniform dilation induced by the SAW, the effective potential can be expressed as \cite{PhysRev.52.688,PhysRev.80.72}
	\begin{equation}
		V\left(\boldsymbol{x},t\right) = -\mathcal{V}e^{-\boldsymbol{\mu}\cdot\left(\boldsymbol{x}-\boldsymbol{C}t\right)}\cos\left[\boldsymbol{q}\cdot\left(\boldsymbol{x}-\boldsymbol{C}t\right)+\theta\right],
	\end{equation}
	where $\tan\theta = -\vert\boldsymbol{\mu}\vert/\vert\boldsymbol{q}\vert$. $\mathcal{V}$ represents the amplitude of the potential \cite{SupplementalMaterials}, and can be tuned experimentally \cite{FootNote1}. {Similar potentials have also been used to predict topological charge pumping \cite{Zheng2016Topological}.}
	
	Under typical experimental conditions, the attenuation length remains small compared to the system dimensions  ($\boldsymbol{\mu} \cdot \boldsymbol{x} \ll 1$), and the long-wavelength approximation for the SAW is valid \cite{campbell2012surface,doi:10.1021/acs.nanolett.4c02941, Wixforth1989Surface, Rotter1998Giant}. Our analysis is restricted to the first order of $\boldsymbol{\mu} \cdot \boldsymbol{x}$. Consider a wave packet centered at $\boldsymbol{x}_c$ at a given time, with its spread much smaller than the SAW wavelength. The Hamiltonian experienced by the wave packet includes the SAW-free Hamiltonian $H_0$ and the effective potential induced by the SAW, $V\left(\boldsymbol{x}, t\right)$, which can be written as
	\begin{equation}
		\hat{H} = \hat{H}_0\left[\hat{\boldsymbol{p}},\hat{\boldsymbol{x}};w\left(\boldsymbol{x}_c\right)\right] + V\left(\boldsymbol{x}_c, t\right),
	\end{equation}
	where $w\left(\boldsymbol{x}_c\right)$ represents other possible mechanical perturbation fields \cite{PhysRevB.59.14915}. Here, we assume $V\left(\boldsymbol{x}, t\right)$ and $w\left(\boldsymbol{x}\right)$ vary slowly on the wave packet scale, allowing to replace $\boldsymbol{x}$ with $\boldsymbol{x}_c$ in $\hat{H}$. 
	
	The non-periodicity of $V\left(\boldsymbol{x}, t\right)$ implies that the eigenstates of $\hat{H}$ are no longer Bloch functions. Nevertheless, the slow attenuation of $V\left(\boldsymbol{x},t\right)$ permits expansion in terms of Wannier functions, where the expansion coefficients are governed by the Wannier-Slater theorem \cite{PhysRev.76.1592}. The expansion coefficients yield bound-state solutions, demonstrating electron trapping in SAW potential troughs, as shown in Fig. \ref{Fig1}(a) (details in Supplemental Materials \cite{SupplementalMaterials}). The eigenstate of $\hat{H}$ can be expressed as 
	\begin{equation}\label{Eq.eigenstate}
		\psi_{\boldsymbol{k}}
		= e^{-iC^2t/2\alpha}\sum_s e^{is\theta} \widetilde{\Psi}\left(\boldsymbol{k}+s\boldsymbol{q}\right) \phi_{\boldsymbol{k}+s\boldsymbol{q}+\boldsymbol{C}/\alpha},
	\end{equation}
	where $\phi_{\boldsymbol{k}}$ is a Bloch wave of $\hat{H}_0$, with $\alpha\equiv1/m^*$ being the inverse effective mass and $s\in\mathbb{Z}$. The folding of Brillouin zone along the SAW propagation direction is manifested through the relation $\psi_{\boldsymbol{k}+\boldsymbol{q}}=e^{-i\theta}\psi_{\boldsymbol{k}}$, where the folding weight $\widetilde{\Psi}(\boldsymbol{k}+s\boldsymbol{q})$ reflects the non-uniformity induced by the electron binding effect (red curve in Fig. \ref{Fig1}(b)) \cite{SupplementalMaterials}. The Gaussian-like folding weight causes $\psi_{\boldsymbol{k}}$ to closely resemble $\phi_{\boldsymbol{k}+\boldsymbol{C}/\alpha}$, indicating stronger SAW binding for electrons with momentum component near $\boldsymbol{k}_\parallel^*=\boldsymbol{C}/\alpha$ along the SAW propagation direction.



	Now we construct a wave packet affected by the SAW centered around $\left(\boldsymbol{x}_c,\boldsymbol{k}_c\right)$, 
	\begin{equation}
		\ket{\Psi\left(\boldsymbol{x}_c,\boldsymbol{k}_c,t\right)} = \int g\left(\boldsymbol{k},t\right)\ket{\psi_{\boldsymbol{k}}},
	\end{equation}	
	where the envelope function $g\left(\boldsymbol{k}, t\right)$ is sharply concentrated in reciprocal space. The momentum of the wave packet is $\boldsymbol{k}_c = \int \vert g\left(\boldsymbol{k}, t\right)\vert^2 \boldsymbol{k}${, where the natural unit system $\hbar=1$ is adopted}. The integration over momentum is performed within the folded Brillouin zone, and the notation $\int $ is shorthand for $ \int \mathrm{d}^2 \boldsymbol{k}/\left(2\pi\right)^2$.

	
	{\it SAW-induced positional shift and dynamics of semiclassical wave packet.---} Although $\psi_{\boldsymbol{k}}$, the eigenstate of the SAW system, does not describe Bloch electrons, the momentum $\boldsymbol{k}$ remains a good quantum number, and charge conservation still holds. Therefore, the center of the wave packet $\ket{\Psi}$ can be defined as $\boldsymbol{x}_c = \braket{\Psi \vert \boldsymbol{x} \vert \Psi}$.
	\begin{equation}
		\boldsymbol{x}_c = \frac{\partial \gamma_c}{\partial \boldsymbol{k}_c} + \sum_s \vert \widetilde{\Psi}\left(\boldsymbol{k}_c+s\boldsymbol{q}\right)\vert^2\boldsymbol{a}_{\boldsymbol{k}_c+s\boldsymbol{q}+\boldsymbol{C}/\alpha}+\frac{\boldsymbol{\tilde{\mu}}}{\sigma}\sqrt{\frac{\pi}{2}},
	\end{equation}
	where $\boldsymbol{\tilde{\mu}}\equiv e^{\boldsymbol{\mu}\cdot\boldsymbol{C}t}\boldsymbol{\mu}$ is the effective attenuation factor, originating from the imaginary part of $\widetilde{\Psi}\left(\boldsymbol{k}\right)$ \cite{SupplementalMaterials}. $\gamma_c\equiv\gamma\left(\boldsymbol{k}_c,t\right)$ is the phase of envelope function $g\left(\boldsymbol{k},t\right)=\vert g\left(\boldsymbol{k},t\right) \vert e^{-i\gamma\left(\boldsymbol{k},t\right)}$ and $\boldsymbol{a}_{\boldsymbol{k}_c}\equiv\braket{u\vert i\nabla_{\boldsymbol{k}_c}u}$ is the Berry connection of $\hat{H}_0$. The second term is the Berry connection, which is folded because of quasi-periodic SAW, with the coefficient $\vert \widetilde{\Psi}\left(\boldsymbol{k}_c+s\boldsymbol{q}\right)\vert^2$. The third term represents the SAW-induced positional shift of the wave packet induced by the attenuated SAW, corresponding to a DC shift current, as confirmed experimentally \cite{10.1063/1.122400,Tang2017Ultra}.
	
	
	The wave packet $\ket{\Psi}$ satisfies the Schrödinger equation, and its Lagrangian can be written as $L = \braket{\Psi \vert i\partial_t - H \vert \Psi}$. Given the sharp envelope function $g\left(\boldsymbol{k}, t\right)$, $L$ can be expressed as a function of both $\boldsymbol{x}_c$ and $\boldsymbol{k}_c$.
	\begin{equation}
		L = -\mathcal{E} + \dot{\boldsymbol{x}_c}\cdot\boldsymbol{k}_c+ \sum_s \vert \widetilde{\Psi}\left(\boldsymbol{k}_c+s\boldsymbol{q}\right)\vert^2 a_{\mathcal{T}} +  \dot{\boldsymbol{k}_c}\cdot \frac{\boldsymbol{\tilde{\mu}}}{q}\sqrt{\frac{\pi}{2}},
	\end{equation}
	where $a_{\mathcal{T}}\equiv\braket{u\vert i\partial_{\mathcal{T}} u}$, with $\partial_{\mathcal{T}}=\partial_t+\dot{\boldsymbol{k}_c}\cdot\nabla_{\boldsymbol{k}_c}+\dot{\boldsymbol{x}_c}\cdot\nabla_{\boldsymbol{x}_c}$ and $q=\vert\boldsymbol{q}\vert$. The center energy $\mathcal{E}$ can be decomposed into on-site energy $\mathcal{E}_0$ and orbital magnetic moment contribution $\Delta \mathcal{E}$, both folded into the new Brillouin zone, as shown in Fig. \ref{Fig1}(b) \cite{SupplementalMaterials}. The last term arises from the SAW-induced positional shift of the wave packet. Electron dynamics in attenuated SAW systems can be expressed as \cite{SupplementalMaterials}
	\begin{subequations}\label{Eq. Dynamics}
		\begin{align}
			\label{Eq. Dyanamic1}&\dot{k}^i_c=-\frac{\partial \mathcal{E}}{\partial {x}^i_c} +   \Omega^F_{x_c^i,\mathcal{T}}+ {\Omega^C_{x_c^i,\mathcal{T}}} ,  \\
			\label{Eq. Dyanamic2}&
			\dot{x}_c^i =+\frac{\partial \mathcal{E}}{\partial {k}^i_c} -\Omega^F_{k_c^i,\mathcal{T}}-{\Omega^C_{k_c^i,\mathcal{T}}}+\frac{\dot{\tilde{\mu}}^i}{q}\sqrt{\frac{\pi}{2}},
		\end{align}
	\end{subequations}
	where $i,j=x,y$. The first terms in Eq.(\ref{Eq. Dynamics}) both represent the folded band contribution. The second terms in Eq.(\ref{Eq. Dynamics}) are the folded Berry curvature contributions,
	\begin{equation}
		\Omega^F_{\lambda,\xi}=\sum_{s}\vert\widetilde{\Psi}\left(\boldsymbol{k}_c+s\boldsymbol{q}\right)\vert^2\left(\braket{\partial_{\lambda}u\vert\partial_{\xi}u} - \braket{\partial_{\xi}u\vert\partial_{\lambda}u}\right),
	\end{equation}
	where $\lambda,\xi=k_c^i,x_c^i,\mathcal{T}$. $\ket{u}$ is the periodic part of Bloch wave of $\hat{H}_0$, being a function of $\boldsymbol{k}_c+s\boldsymbol{q}+\boldsymbol{C}/\alpha$. These terms resemble the Berry curvatures in conventional semiclassical theory \cite{PhysRevB.59.14915}, but differ fundamentally due to SAW-induced Brillouin zone folding, leading to substantial modifications in the electron dynamics.
	

	The third terms in Eq. (\ref{Eq. Dynamics}), which we call the {curvature cross terms}, take the form
	\begin{equation}
		{\Omega^C_{\lambda,\xi}} = \sum_{s}\frac{\partial}{\partial \lambda}\vert\widetilde{\Psi}\left(\boldsymbol{k}_c+s\boldsymbol{q}\right)\vert^2 a_{\xi}-\left(\lambda\leftrightarrow\xi\right),
	\end{equation}
	where $a_{\xi}$ are Berry connections of $\hat{H}_0$. The curvature cross terms emerge exclusively in electron dynamics when accounting for SAW-induced non-uniform Brillouin zone folding, having no counterpart in unperturbed systems. {These terms can be rigorously proven to be gauge-independent \cite{SupplementalMaterials}.} {We note that both the folded Berry curvature $\Omega^F_{\lambda,\xi}$ and the curvature cross term $\Omega^C_{\lambda,\xi}$ stem from the same physical origin. Mathematically, both terms are recast into a form that matches the original electron dynamics (see End Matter for details).}
	
	
	

	{\it Nontrivial electric and thermal Response.---} For simplicity, we now consider a spatially periodic, time-independent Hamiltonian $\hat{H}_0$ that depends only on the momentum $\boldsymbol{k}$. As a result, the Berry curvatures and Berry connections associated with $\boldsymbol{x}$ and $t$ vanish. We designate the transport direction of the SAW as $x$-axis and the vertical direction as $y$-axis. The electron dynamics Eq.(\ref{Eq. Dynamics}) can be simplified as
	\begin{subequations}\label{Acousto_EOM}
		\begin{align}
			&\label{Acousto_EOMa}\dot{\boldsymbol{k}}_c=-\mathcal{J}_a \boldsymbol{s}e^{-\boldsymbol{\mu}\cdot\left(\boldsymbol{x}_c-\boldsymbol{C}t\right)}\sin\left[\boldsymbol{q}\cdot \left(\boldsymbol{x}_c-\boldsymbol{C}t\right)\right], \\
			&\label{Acousto_EOMb}
			\dot{\boldsymbol{x}}_c = \frac{\partial \mathcal{E}}{\partial {\boldsymbol{k}}_c} -\dot{\boldsymbol{k}}_c\times\left(\boldsymbol{\Omega}^{\boldsymbol{F}}_{\boldsymbol{k}_c}+\boldsymbol{\Omega}^{\boldsymbol{C}}_{\boldsymbol{k}_c}\right)+\frac{\boldsymbol{\dot{\tilde{\mu}}}}{q}\sqrt{\frac{\pi}{2}},		
		\end{align}
	\end{subequations}
	where $\mathcal{J}_a = \mathcal{V}\sqrt{\boldsymbol{\mu}^2+\boldsymbol{q}^2}/\vert\boldsymbol{s}\vert $ \cite{SupplementalMaterials}. We define the folded Berry curvature vector $\boldsymbol{\Omega}^{\boldsymbol{F}}_{\boldsymbol{k}}\equiv\left[0,0,\Omega^F_{k_xk_y}\right]$ and the curvature cross term vector $\boldsymbol{\Omega}^{\boldsymbol{C}}_{\boldsymbol{k}}\equiv\left[0,0,\Omega^C_{k_xk_y}\right]$ here for simplicity. Consequently, the acousto-electric Hall current derived from semiclassical dynamical theory is
	\begin{equation}
		\boldsymbol{J}^{e}_{Hall}={e}\mathcal{J}_a\boldsymbol{u}\times\int f\left(\boldsymbol{k},T\right)\left(\boldsymbol{\Omega}^{\boldsymbol{F}}_{\boldsymbol{k}}+\boldsymbol{\Omega}^{\boldsymbol{C}}_{\boldsymbol{k}}\right),
	\end{equation}
	where $f(\boldsymbol{k},T)$ represents the Fermi-Dirac distribution for the folded band structure of $\hat{H}_0$. {Symmetry analysis confirms that the SAW breaks time-reversal and spatial-inversion symmetries, allowing a finite Hall current $\boldsymbol{J}^{e}_{Hall}$ to emerge even with a time-reversal symmetric system $\hat{H}_0$ \cite{SupplementalMaterials}. This mechanism differs from the conventional second-order DC acousto-electric response, which generally requires additional breaking of spatial symmetries within $\hat{H}_0$.} The electron dynamics additionally yield a DC current parallel to the SAW propagation direction, given by
	\begin{equation}
		\boldsymbol{J}^{e}_{drag}=-e\sqrt{\frac{\pi}{2}} \frac{\boldsymbol{\dot{\tilde{\mu}}}}{q} \braket{n\left(T\right)},
	\end{equation}
	where $\braket{n(T)}\equiv\int f\left(\boldsymbol{k},T\right)$ denotes the averaged electron occupation number at temperature $T$. This current corresponds to the SAW-induced drag DC current, which is proportional to the decay rate $\boldsymbol{\mu}$, in agreement with both experimental observations \cite{PhysRevLett.128.256601} and our theoretical description of SAW-driven electron dynamics. Equation (\ref{Eq. Dyanamic2}) elucidates the microscopic mechanism of the drag DC current, which stems from the positional shift of the electron wave packet induced by the SAW-mediated breaking of periodicity.
	

	
	
	
	However, the SAW frequency (on the order of MHz) lies well beyond the detectable range of conventional phase-sensitive measurements using lock-in amplifiers. Additionally, due to the form of $\boldsymbol{u}$, the time-averaged current vanishes. Consequently, the first-order acousto-electric Hall effect may prove challenging to observe with current experimental techniques.
	
	To experimentally determine the response coefficient of the acousto-electric Hall effect, a static or low-frequency electric field $\boldsymbol{E}$ is introduced, resulting Eq.(\ref{Acousto_EOMa}) to $\dot{\boldsymbol{k}}_c=-\mathcal{J}_a\boldsymbol{u}-e\boldsymbol{E}$. This configuration gives rise to an additional electric current, perpendicular to $\boldsymbol{E}$, oscillating at the same frequency as $\boldsymbol{E}$. The corresponding Hall conductivity can be expressed as
	\begin{equation}
		\sigma_{xy} = {e^2}\int f\left(\boldsymbol{k},T\right)\left(\Omega^F_{k_xk_y}+\Omega^C_{k_xk_y}\right),
	\end{equation}	
	which is identical to the acousto-electric Hall coefficient up to a constant multiplier. Consequently, even in a topologically trivial system where the Chern number vanishes, the response function $\sigma_{xy}$ may not necessarily vanish. Importantly, this nontrivial response emerges from treating SAW as a modulation of elecron dynamics, beyond the conventional electric field approximation. Moreover, when the input direction of SAW is rotated relative to the system, the response function may exhibit an angular dependence, which typically shares the same symmetry as the system.
	
	
	In addition to applying an electric field, a temperature gradient $\Delta T$ can also be introduced to probe the Brillouin zone folding effect induced by SAW. The temperature gradient may generate a thermal current and electrical current, known as the anomalous thermal Hall effect \cite{PhysRevLett.107.236601,PhysRevB.101.235430} and anomalous Nernst effect \cite{PhysRevLett.118.136601}, whose response function can be expressed as
	\begin{equation}
		\begin{pmatrix}
			\kappa_{xy}\\ \alpha_{xy}
		\end{pmatrix}
		= \frac{1}{T}\int\int_{\mathcal{E}}^{\infty} d\eta 
		\begin{pmatrix}
			-\left(\eta-\mu\right)^2\\ e\left(\eta-\mu\right)
		\end{pmatrix}
		f'\left(\Omega^F_{k_xk_y}+\Omega^C_{k_xk_y}\right),
	\end{equation}
	where $f'\left(\eta\right)=\partial f/\partial\eta$ represents the derivative of the Fermi-Dirac distribution with respect to energy $\eta$, and $\mu$ denotes the chemical potential. Even in a topological trivial system, $\kappa_{xy}$ and $\alpha_{xy}$ do not vanish, and the Wiedemann-Franz law $\kappa_{xy}^0 = -\left(\pi^2k_B^2T/3e^2\right) \sigma_{xy}^0$ and Mott relation $\alpha_{xy}^0 = \left(\pi^2k_B^2T/3e\right) \partial\sigma_{xy}^{0}/\partial\mu$ can be demonstrated to hold at low-temperature regime by employing the Sommerfeld expansion \cite{SupplementalMaterials}. These nontrivial electric and thermal transport responses and their relationship provide us more possibility to confirm our semiclassical electron dynamics in SAW system.
	
	
	{\it Angular-Dependent Acousto-Electric Hall Effect.---} The anomalous velocity of electron in SAW systems acquires a directional dependence on the SAW propagation direction due to the non-uniform Brillouin zone folding. In the presence of the lattice, the continuous rotational symmetry of a uniform 2D system is broken down to the discrete point group symmetry of the crystal, leading to angular dependence of the response function in the acousto-electric Hall effect. Moreover, the non-uniform Brillouin zone folding induces momentum-dependent variations in the Berry curvature contribution. Consequently, even in topologically trivial systems with vanishing Chern number and absent quantum anomalous Hall effect, the angular-dependent acousto-electric Hall effect persists. We apply the semiclassical theory to AB stacked bilayer graphene \cite{PhysRevLett.107.256801,PhysRevB.81.125435,PhysRevB.74.165310,zan2024observationelectricalhighharmonicgeneration} and group-VIB transition metal dichalcogenides $\mathrm{MX_2}$ ($\mathrm{M=Mo,W}$;$\mathrm{X=S,Se,Te}$) \cite{app6100284,PhysRevB.88.085433,PhysRevB.109.245412}, two systems with different topological structures.

	In undoped bilayer graphene, spontaneous magnetization may occur, giving rise to a small exchange field and spontaneously broken time-reversal symmetry \cite{doi:10.1126/science.abm8386,RN61,RN62}. We therefore investigate the combined effects of this spontaneous time-reversal symmetry breaking and Rashba-type spin-orbit coupling (SOC). Then, the Hamiltonian can be written as $\hat{H}=\hat{H}_0+\hat{H}'$.
	\begin{equation}
		\hat{H}_0=
		\begin{pmatrix}
			\mathcal{E}_{A1}-\Delta/2&-\gamma_0h\left(\boldsymbol{k}\right) & \gamma_4h\left(\boldsymbol{k}\right)& -\gamma_3h^*\left(\boldsymbol{k}\right)\\
			-\gamma_0h^*\left(\boldsymbol{k}\right) &\mathcal{E}_{B1}-\Delta/2& \gamma_1&\gamma_4h\left(\boldsymbol{k}\right)\\
			\gamma_4h^*\left(\boldsymbol{k}\right)&\gamma_1&\mathcal{E}_{A2}+\Delta/2&-\gamma_0h\left(\boldsymbol{k}\right)\\
			-\gamma_3h\left(\boldsymbol{k}\right) & \gamma_4h^*\left(\boldsymbol{k}\right) & 	-\gamma_0h^*\left(\boldsymbol{k}\right) & \mathcal{E}_{B2}+\Delta/2
		\end{pmatrix}
	\end{equation}
	and $\hat{H}' = i\lambda_R\boldsymbol{\hat{e}_z}\cdot\left(\boldsymbol{d}\times\boldsymbol{\sigma}\right)+M\sigma_z$, where $h\left(\boldsymbol{k}\right)=e^{ik_ya/\sqrt{3}}+2e^{-ik_ya/2\sqrt{3}}\cos\left(k_xa/2\right)$. Let $a$ be the lattice constant, $\Delta$ denotes the interlayer potential difference induced by the displacement field, $\boldsymbol{d}$ represents the nearest-neighbor bond vectors, and $M$ is the magnetization strength \cite{zan2024observationelectricalhighharmonicgeneration,SupplementalMaterials}.
	
	For group-VIB transition metal dichalcogenides $\mathrm{MX_2}$ (M = Mo, W; X = S, Se, Te), we employ a six-band tight-binding Hamiltonian derived from first-principles calculations \cite{app6100284,PhysRevB.88.085433,PhysRevB.109.245412}, with the explicit form
	\begin{equation}
		\mathcal{H} = 
		\begin{pmatrix}
			\mathcal{E}_M + 2 t_i^{MM}\cos\left(\boldsymbol{k}\cdot\boldsymbol{a}_i\right) & t_i^{MX}e^{-i\boldsymbol{k}\cdot\boldsymbol{\delta}_i } \\
			t_i^{MX}e^{+i\boldsymbol{k}\cdot\boldsymbol{\delta}_i } & \mathcal{E}_X + 2 t_i^{XX}\cos\left(\boldsymbol{k}\cdot\boldsymbol{a}_i\right)
		\end{pmatrix},
	\end{equation}
	where $\boldsymbol{a}_i$ represents the primitive vector, and $\boldsymbol{\delta}_i $ denotes the vector pointing from an A site to a B site in honeycomb lattice \cite{app6100284} (details in supplymental materials \cite{SupplementalMaterials}). 
	
	
	\begin{figure}[t]
		\centering
		\includegraphics[width=1\linewidth]{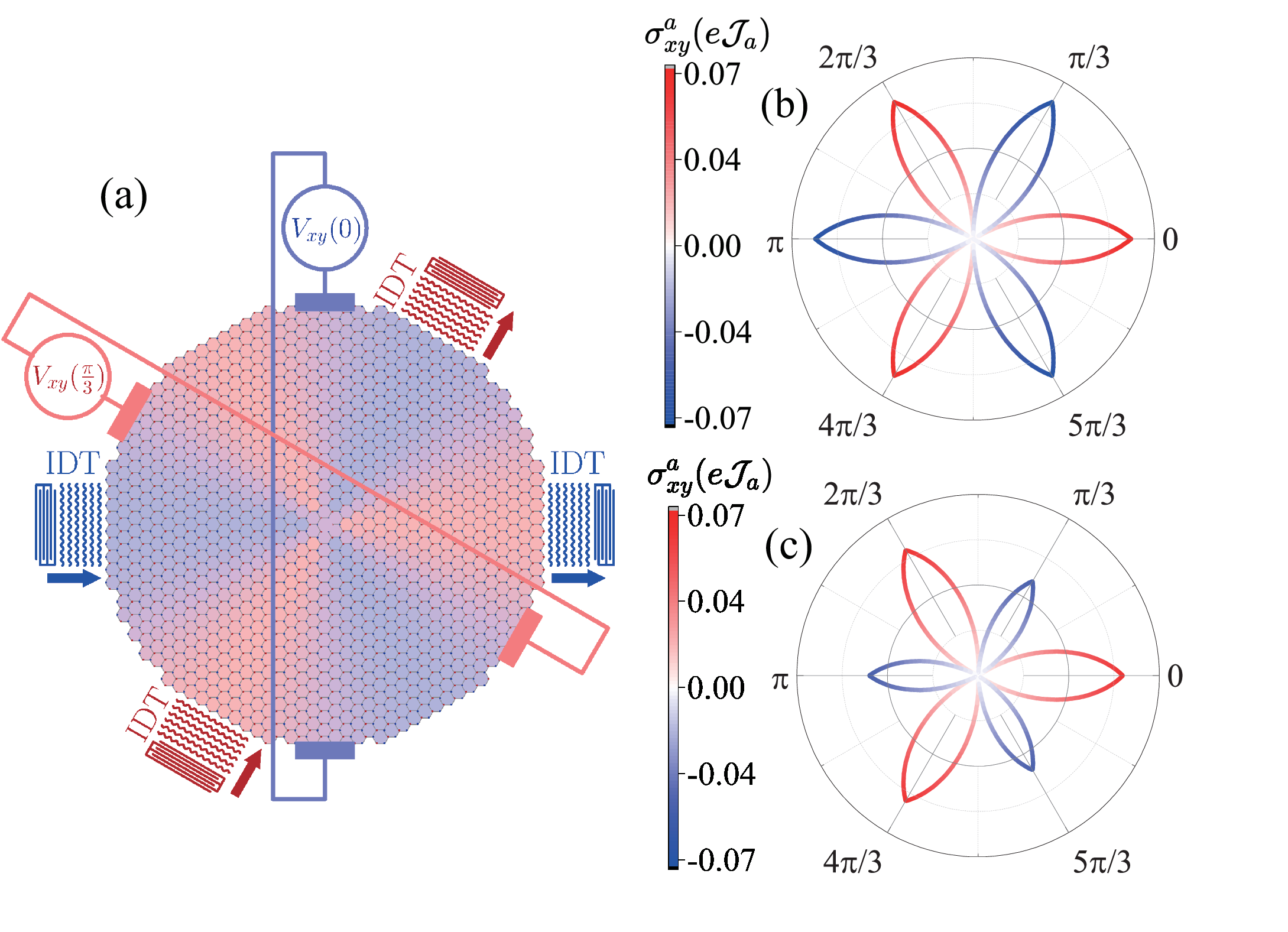}
		\caption{(a) Measurement setup for the angular-dependent acousto-electric Hall effect using interdigital transducers (IDTs) to generate SAW, with atomic sites A (red dots) and B (blue dots) explicitly labeled. The SAW propagation direction is defined as 0 radians along the armchair direction (A$\rightarrow$B, blue IDTs) and $\pi/3$ along an inequivalent armchair direction (B$\rightarrow$A, red IDTs).
		The angular dependence of the acousto-electric Hall conductivity $\sigma^a_{xy}$ is shown for (b) bilayer graphene and (c) $\mathrm{MX_2}$ systems. The calculations adopt the following SAW parameters: phase velocity $C=3000$m/s, frequency $\omega=500$MHz and input power $P_0=0.001\mu$W.}
		\label{Fig2}
	\end{figure}

	The angular-dependent acousto-electric Hall effect persists in both systems despite their zero Chern numbers and absent quantum anomalous Hall effects (as shown in Fig. \ref{Fig2}(b)-(c)), and can be experimentally measured using the setup depicted in Fig. \ref{Fig2}(a). Due to the $C_3$ symmetry, $\sigma_{xy}^a\left(0\right)$ and $\sigma_{xy}^a\left(\pi/3\right)$ may differ, resulting in a six-petal angular pattern for the acousto-electric Hall conductivity. In bilayer graphene, the Berry curvature exhibits strong localization at K/K' valleys, creating Berry curvature hot spots \cite{zan2024observationelectricalhighharmonicgeneration}. This localization causes SAW propagating along $0$ and $\pi/3$ orientations to possibly probe the complete topological signatures of hot spots, yielding conductivity values that are precisely opposite. Conversely, in $\mathrm{MX_2}$ systems, the Berry curvature displays broader spatial distribution. As a result, SAW at $0$ and $\pi/3$ orientations intercept only partial topological characteristics, resulting in conductivity values that are not perfectly opposite \cite{SupplementalMaterials}. The angular-dependent acousto-electric Hall effect not only reflects the system's intrinsic symmetry but also reveals the Berry curvature distribution through angle-resolved conductivity variations, a capability unattainable in conventional transport measurements. {More numerical calculations for both systems are provided in Supplymentary Materials \cite{SupplementalMaterials}.}
	

	{\it Discussion and Conclusion.---} In summary, we have derived the semiclassical electron dynamics in the presence of SAW and demonstrated the effects of the folded Berry curvature and curvature cross terms, originating from the non-uniform Brillouin zone folding effect. Our theory provides a microscopic mechanism for the longitudinal DC current proportional to the attenuation rate $\boldsymbol{\mu}$, and predicts the emergence of acousto-electric Hall effect even in topologically trivial systems. The SAW-induced non-uniform folding effect generates unique modifications to charge and heat transport, manifesting in anomalous Hall, thermal Hall, and Nernst effect signatures. The angular-dependent acousto-electric Hall effect is demonstrated in both bilayer graphene and group-VIB transition metal dichalcogenides. This effect can reveal the Berry curvature distribution by varying the SAW propagation direction, which cannot be achieved by conventional transport measurements. Our work establishes a new framework for studying electron dynamics and the transport properties in quasi-periodic external fields. 
	
	{In addition to the linear acousto-electric Hall effect discussed above, our theory also modifies the second-order acousto-electric response. The derived second-order DC response differs characteristically from conventional predictions due to the non-uniform folding effect, with elementary derivations provided in the Supplementary Materials \cite{SupplementalMaterials}. Furthermore, our theory describes the SAW-induced electron binding via an effective potential that incorporates the polaron concept. The omission of electron backaction may lead to quantitative differences from a rigorous polaron theory, but does not affect the validity of the physical picture we establish \cite{frohlich1950xx,frohlich1954electrons}. Our semiclassical framework focuses on the intrinsic mechanisms of SAW-modulated electron dynamics, as intrinsic contributions play a decisive role in the family of Hall effects. Moreover, all parameters in our general theory can be directly obtained from material-specific first-principles calculations. It should be noted that extrinsic effects, such as the disorder-induced side-jump and skew scattering mechanism, also significantly influence Hall transport. These effects constitute an important direction for future investigation.}
	
	{\it Acknowledge.---} We thank Shuguang Cheng, Pengyi Liu, Wen-Bo Dai and Humian Zhou for illuminating discussions.
	This work was financially supported by National Key R and D Program of China (Grants No. 2024YFA1409003, and 2022YFA1403700), NSFC (Grants Nos. 12350401), and X.C.X. acknowledges the support from the Innovation Program for Quantum Science and Technology (Grant No. 2021ZD0302400).

	\bibliography{References}

\begin{thebibliography}{75}
\bibitem [{\citenamefont {{Here}}()}]{FootNote1}%
  \BibitemOpen
  \bibfield  {author} {\bibinfo {author} {\bibnamefont {{Here}}},\ }\href@noop
  {} {}\bibinfo {note} {{we consider a time-dependent excitation $\mathcal{V}
  e^{\boldsymbol{\mu} \cdot \boldsymbol{C} t}$, which is both experimentally
  tunable and analytically tractable. The expressions for $\sigma_{xy}$,
  $\kappa_{xy}$, and $\alpha_{xy}$ remain valid even if the excitation is
  replaced by a static $\mathcal{V}$.}}\BibitemShut {Stop}%
\end{thebibliography}
	
	{
	\section*{End Matter}
	
	The terms $\Omega^F_{\lambda,\xi}$ and $\Omega^C_{\lambda,\xi}$ in the electron dynamics (Eq. \ref{Eq. Dynamics}) share the same physical origin, making it impossible to separately measure their contributions in transport. This homology becomes evident when defining a modified wave function $\ket{\tilde{u}_{\boldsymbol{k}}}$, which allows for the construction of a modified Berry connection and the corresponding Berry curvature. $\ket{\tilde{u}_{\boldsymbol{k}}}$ is defined as
	\begin{equation}
		\ket{\tilde{u}_{\boldsymbol{k}}} = e^{-i\boldsymbol{k}\cdot\boldsymbol{x}}\ket{\psi_{\boldsymbol{k}}}=\sum_s e^{is\theta} \widetilde{\Psi}\left(\boldsymbol{k}+s\boldsymbol{q}\right) \ket{u_{\boldsymbol{k}+s\boldsymbol{q}+\boldsymbol{C}/\alpha}}.
	\end{equation}
	The modulated Berry connection can be calculated  as
	\begin{equation}
		\begin{aligned}
			\tilde{a}_{\mathcal{T}}=&\braket{\tilde{u}_{\boldsymbol{k}_c}\vert i\partial_{\mathcal{T}}\vert\tilde{u}_{\boldsymbol{k}_c}}
			= \dot{\boldsymbol{k}_c}\cdot \frac{\boldsymbol{\tilde{\mu}}}{\boldsymbol{q}}\sqrt{\frac{\pi}{2}}\\
			&+\sum_{s}\vert \widetilde{\Psi}\left(\boldsymbol{k}+s\boldsymbol{q}\right)\vert^2\braket{u_{\boldsymbol{k}+s\boldsymbol{q}+\boldsymbol{C}/\alpha}\vert i\partial_{\mathcal{T}}\vert u_{\boldsymbol{k}+s\boldsymbol{q}+\boldsymbol{C}/\alpha}},
		\end{aligned}
	\end{equation}
	and therefore the Lagragian can then be written as
	\begin{equation}
		L = -\mathcal{E} + \dot{\boldsymbol{x}_c}\cdot\boldsymbol{k}_c + \braket{\tilde{u}_{\boldsymbol{k}_c}\vert i\partial_{\mathcal{T}}\vert\tilde{u}_{\boldsymbol{k}_c}}.
	\end{equation}
	The modulated Berry curvature is
	\begin{subequations}
		\begin{align}
			\widetilde{\Omega}_{x^i_c,\mathcal{T}} =& \Omega^F_{x^i_c,\mathcal{T}} + \Omega^C_{x^i_c,\mathcal{T}},\\
			\widetilde{\Omega}_{k^i_c,\mathcal{T}} =& \Omega^F_{k^i_c,\mathcal{T}} + \Omega^C_{k^i_c,\mathcal{T}} - \frac{\dot{\tilde{\mu}}^i}{q}\sqrt{\frac{\pi}{2}},
		\end{align}
	\end{subequations}
	As a result, the electron dynamics can be written as 
	\begin{subequations}
		\begin{align}
			&\dot{k}^i_c=-\frac{\partial \mathcal{E}}{\partial {x}^i_c} + \widetilde{\Omega}_{x^i_c,\mathcal{T}},  \\
			&
			\dot{x}_c^i =+\frac{\partial \mathcal{E}}{\partial {k}^i_c} -\widetilde{\Omega}_{k^i_c,\mathcal{T}}.
		\end{align}
	\end{subequations}
	The mathematical equivalence between the SAW-modulated electron dynamics and the original semiclassical equations is established through modified wave function $\ket{\tilde{u}_{\boldsymbol{k}}}$, demonstrating both the homology between $\Omega^F_{\lambda,\xi}$ and $\Omega^C_{\lambda,\xi}$ and the consistency of our theoretical framework. We highlight the $\Omega^C_{\lambda,\xi}$ terms separately to emphasize the distinctive role of the SAW-induced non-uniform Brillouin zone folding effect. Detailed derivations are provided in the Supplementary Material \cite{SupplementalMaterials}.
	
	In transport measurements, $\Omega^F_{\lambda,\xi}$ and $\Omega^C_{\lambda,\xi}$ cannot be measured separately. Isolating the contribution of the curvature cross term $\Omega^C_{\lambda,\xi}$ requires systems with minimal $\Omega^F_{\lambda,\xi}$. In bilayer graphene, Berry curvature hot spots are concentrated near the $K$ points, while the folding function $\vert\widetilde{\Psi}\vert^2$ is predominantly distributed around $\Gamma$, resulting in a negligible $\Omega^F_{\lambda,\xi}$ and making $\sigma^a_{xy}$ dominated by $\Omega^C_{\lambda,\xi}$. Future studies aiming to probe the $\Omega^C_{\lambda,\xi}$ should therefore seek systems where Berry curvature is strongly localized and spatially separated from the distribution of $\vert\widetilde{\Psi}\vert^2$. For instance, in hexagonal lattices where Berry curvature peaks near $K$, systems with small electron effective mass are preferred as they yield smaller $C/\alpha$ values, thus suppressing the $\Omega^F_{\lambda,\xi}$ contribution to transport.
	
}
	
\end{document}